
\input harvmac
\baselineskip8pt
\Title{\vbox
{\baselineskip 6pt{\hbox{ }}{\hbox
{Imperial/TP/94-95/35 }}{\hbox{hep-th/9505145}} {\hbox{ }}} }
{\vbox{\centerline { Closed strings }
\centerline { in uniform  magnetic field backgrounds  }
}}

\vskip -5 true pt

\centerline{   A.A. Tseytlin\footnote{$^{\star}$}{\baselineskip8pt
e-mail address: tseytlin@ic.ac.uk}\footnote{$^{\dagger}$}{\baselineskip6pt
On leave  from Lebedev  Physics
Institute, Moscow, Russia.} }

\smallskip\smallskip
\centerline {\it  Theoretical Physics Group, Blackett Laboratory}
\smallskip

\centerline {\it  Imperial College,  London SW7 2BZ, U.K. }
\bigskip
\centerline {\bf Abstract}
\medskip
\baselineskip8pt
\noindent
We consider a class of conformal models  describing closed
strings in  axially symmetric stationary magnetic flux tube
backgrounds. These models are closed string analogs of the
Landau model of a particle in a magnetic field or the model
of an  open string in a constant magnetic field.
They are  interesting  examples of solvable unitary conformal
string theories  with  non-trivial 4-dimensional  curved
space-time interpretation. In particular, their quantum
Hamiltonian can be expressed in terms of free fields
and the physical spectrum and string  partition function can be
explicitly determined. In addition to the presence of tachyonic
instabilities and existence of critical values of magnetic field
the closed string spectrum exhibits  also some novel
features which were  absent in the open string case.

\bigskip\bigskip
\bigskip\bigskip

{\it  Contribution to  Proceedings of ``Strings 95",  USC, Los Angeles, March
13-18, 1995 }

\Date {May  1995}

\noblackbox
\baselineskip 14pt plus 2pt minus 2pt
\vfill\eject

\def \eq#1 {\eqno{(#1)}}
\def \ha {{1\over 2}}

\def \ov  {\over  }
\def\np  { {\it  Nucl. Phys. }}
\def \pl { {\it  Phys. Lett. }}
\def \mpl{ {\it  Mod. Phys. Lett. }}
\def \prl{ {\it  Phys. Rev. Lett. }}
\def \pr { {\it  Phys. Rev. }}

 \def \ijmp{ {\it  Int. J. Mod. Phys. }}
\def \n {\nu}
\def \cos {\ {\cos} \ }\def \htt { {\hat t}}
\def \h {\hat }

\def \l {\lambda}
\def \p {\phi}
\def \vp {\varphi}
\def  \g {\gamma}

\def \r {\rho}

\def \s {\sigma}
\def \t {\tau}
\def \b {\beta}

\def \g {\gamma}

\def \e#1 {{ e}^{#1}}
\def \bd {\bar \partial}
\def \del {\partial}
\def \four {{\textstyle {1\over 4}}}
\def \td {\tilde}
\def \tvp {\tilde \varphi}
\def \a {\alpha}

\def \m {\mu}

\def \H {{\cal H}}
\def \vp {\varphi}
\def \ep {\epsilon}
\def \inv {^{-1}}
\def \ha {{\textstyle{1 \over 2}}}
\def \ov {\over }
\def \te {\theta}
\def \sm {$\s$-model\ }
\def \sms {$\s$-models\ }

\leftline{\tenbf 1. Introduction}
\vglue0.4cm
The study of behavior of systems of particles and fields
in  an external uniform magnetic field is one of the basic
problems in theoretical physics and has long history.
Like temperature,  the magnetic  field plays the role
of a probe  which  may  be used to reveal
various  properties of a  system.
A remarkable feature  of the uniform magnetic field problems
in quantum mechanics and QED  is their  solvability$^{1,2}$.
That  applies to certain extent
also to   gauge theories in  abelian
magnetic environment
various aspects of which  (possibility of restoration of broken symmetry,
instability of  magnetic background in non-abelian  models, formation of vacuum
condensates, etc.)
were  extensively  studied in the past$^{3-6}$.

It is  interesting  to try to address a  similar problem
in the context of string theory (with one of the standard motivations  that
this may eventually help  us to
learn about its possible phase structure).
The reason why the  quantum-mechanical or field-theoretic problem of  a
particle in  a  uniform abelian
(electro)magnetic field is exactly solvable
 is that the action  $I= \int d\t  [\dot x^\m \dot x^\m   +i
 \dot x^\m A_\m (x)]$  (which determines the Hamiltonian in quantum mechanics
and the  heat kernel in field theory)
becomes gaussian if the field  strength is constant,
 $$A_\m =-\ha F_{\m\n} x^\n\ ,   \ \ \ F_{\m\n} = const \ . $$
The same is true also in (abelian) open string theory
where the interaction takes place  only at the boundary points
$$I=  {1\ov 4\pi \a'} \int d^2 \s \  \del_a x^\m \del^a x^\m  +  i \int d\tau \
A_\m (x) \dot x^\m\ ,  $$
 and
thus the resulting   gaussian path integral can be  computed exactly$^7$.
This is a consistent `on-shell' problem  since $F_{\m\n}=const$ is
an exact  solution of the effective field
equations$^7$ of the open string theory.
Indeed, the  corresponding 2d world-sheet theory  represents a conformal field
theory$^8$
which can be solved
explicitly in terms of free oscillators  thus representing  a
generalization  of the   Landau  problem   in quantum particle mechanics.
As a result, one is able to  determine the spectrum of an open  string moving
in a constant magnetic field$^{8,9,10}$.

A novel  feature of this spectrum as compared to the free string spectrum
 is the presence of  new tachyonic states  above certain critical values of the
magnetic field$^{8,10}$.  That implies that
 the constant magnetic field background  is unstable
 in the open string theory as it is in the  non-abelian
gauge theory$^5$. The qualitative reason for this instability is that the
 free  open string  contains  electrically  charged higher spin
massive particle states. The latter are expected to  have  (approximately)
 the Landau spectrum
$$M^2 = M^2_0 + Q {\cal H}( 2l+1- g S)\ , $$
where $Q$ is the charge (the same for all open string states),
$ {\cal H}$ is magnetic field in $x_3$ direction, $g$ is a gyromagnetic ratio
(the effective weak field value of $g$  is
2  for the non-minimally coupled higher spin  open string states$^{11}$), $S$
is the $x_3$-component of the spin and $l=0,1,2,...,$ is the Landau level.
Thus $M^2$ can become negative for large enough values of $\H$,
i.e., $\H > \H_{cr}= M^2_0/Q$ for spin 1 charged states.
That applies, for example, to  $W$-bosons in the context of electroweak
theory$^6$
suggesting the presence of a  transition to a phase with a $W$-condensate
(at higher critical field where Higgs field becomes massless the full
electroweak symmetry is restored$^6$).
Note that in the case of unbroken gauge theory with massless
charged vector particles  the instability is present for any (e.g.,
infinitesimal) value of the magnetic field$^5$. Such `infinitesimal'
instability is thus to be expected   in  the
open string theory with {\it non}-abelian Chan-Paton symmetry
(where the constant magnetic field problem  does not appear to be easily
solvable)  and in {\it closed} string theory (discussed  below).

The
infra-red  instability of a magnetic background is not cured
 by supersymmetry, i.e.
it remains also in supersymmetric gauge theories
(e.g.,  in ultra-violet finite $N=4$ supersymmetric
  Yang-Mills theory$^{12}$)
since the small fluctuation operator for the gauge field $-\delta_{\m\n} D^2 -
2 F_{\m\n} $
still has negative modes
  due to the `anomalous magnetic moment' term.
This is not surprising  given that
 the magnetic field breaks  Lorentz invariance and
supersymmetry.   This instability
  is indeed  present  in the  Neveu-Schwarz  sector
of the open superstring theory$^{ 10 }$ (the fermionic Ramond states remain
non-tachyonic as in field theory).

Assuming  that  it is  important  to try to generalize the
open string results to the case of  realistic closed
models (see,  e.g., ref.{13})
the main  question$^7$, however,
 is whether the uniform magnetic field problem is  actually tractable
in  {\it closed}  string theories.
An apparent   answer is `no' since  the abelian vector  field  must now  be
coupled to the  internal points of the string
and such interaction terms, e.g.,
$$L =  \del_a y \del^a y +  A_\m (x) \del_a x^\m \del^a y + ... \ , $$
in bosonic string
or type II superstring ($y$ is a compact internal Kaluza-Klein field
that `charges'  the string),
or
$$L = \bar \psi \gamma^a [ \del_a +  A_\m (x)\del_a x^\m ] \psi  + ... \ , $$
in the heterotic string,
do not become gaussian  for $A_\m =-\ha F_{\m\n} x^\n$.

One should note, however, that in contrast to the  tree level
abelian
open string case, the  $F_{\m\n}=const$ background in flat space does not
represent a solution of  a  {\it closed} string theory,
i.e. the above interaction terms  added to the  free string Lagrangians
do not give  conformally invariant 2d  $\s$-models.
Indeed, since the closed string  theory contains gravity,
 a uniform  magnetic field which has a non-vanishing energy
must curve the space (as well as
possibly induce other `massless' backgrounds).
One should thus first find a consistent  conformal model
which is a closed string analog
of the  uniform magnetic field background  in
the flat space  field (or open string) theory
and {\it then} address the question of its solvability.
Remarkably, it turns out$^{14-16}$ that extra terms  which
should be added  to   the above  closed string actions in order to  satisfy
 the conformal invariance condition (i.e. to satisfy the closed string
effective field  equations)
produce  exactly solvable  2d models!

In order to  try construct   conformal
$\s$-models  which can be interpreted as describing  closed string
in a uniform magnetic field background
it is  useful
 to look at  possible  `magnetic'  solutions
of low-energy effective string equations.
There  is a   simple  analogue  of a uniform
  magnetic  field background
  in  the Einstein-Maxwell
theory:  the static cylindrically symmetric
 Melvin `magnetic universe' or `magnetic flux tube'  solution$^{17}$.
It has $R^4$  topology  and   can be  considered$^{19}$   as
a gravitational analog of the Abrikosov-Nielsen-Olesen vortex$^{18}$
with the magnetic pressure (due to  repulsion of Faraday's flux lines)
being balanced not by  Higgs field but by gravitational attraction.
The magnetic field is  approximately constant inside  the
tube and decays  to zero  at infinity in the direction orthogonal to
$x_3$-axis.
 Several interesting  features  of the Melvin  solution in the context of
Kaluza-Klein (super)gravity (e.g.,  instability against monopole  or  magnetic
black hole pair creation)  were discussed  ref.19 (see also ref.{20}).
This  Einstein-Maxwell (`$a=0$' Melvin) solution
has  two straightforward  analogs$^{21}$  among solutions of
 low-energy closed string  theory (heterotic string or $D>4 $  bosonic string
or superstring toroidally
compactified to $D=4$). In what follows we shall mostly consider the case
when the magnetic field has Kaluza-Klein origin.
Assuming $x^5=y$ is a compact internal coordinate,
 the   $D=5$ string effective
action  can be expressed in terms of $D=4$ fields:
 metric $G_{\m\n}$, dilaton $\p$,  antisymmetric tensor
 $B_{\m\n}$, {\it  two}  vector fields ${\cal A}_\m$ and ${\cal B}_\m$ (related
to $G_{5\m} $ and $B_{5\m}$)   and
the `modulus'  $\s$. The  dilatonic (`$a=1$')  and  Kaluza-Klein (`$a=\sqrt
3$')
Melvin solutions have zero
$B_{\m\n}$  but  $\p$ or $\s$ being non-constant.

In addition to  the Melvin solutions,  the string theory  equations admits also
 { another}
  natural uniform magnetic field  solution$^{22,14} $  which
has $B_{\m\n}\not=0$ and thus has no  counterpart  in the Einstein-Maxwell
theory.
It   can be considered as a direct  closed string analog
of the $F_{\m\n}=const$ solution of the Maxwell theory
since here the magnetic field is  indeed {\it constant } (and covariantly
constant)  throughout the space
(dilaton is constant as well).
Its metric $ds^2= - (dt + A_i dx^i)^2 + dx_idx^i + dx_3^2, \ \ A_i=
-\ha F_{ij}
x^j, \ \  (i=1,2$)
is that of a product of a real line $R$ and the  Heisenberg group space
$H_3$,  and  the antisymmetric tensor field strength
 is equal to the  constant magnetic field
$H_{tij}= F_{ij}=const $.
\footnote{$^a  $}{\baselineskip6pt
There are  also other non-uniform magnetic monopole type  string backgrounds
which  will not  be discuss here (see, in particular, ref.23).
In addition to the $a=0$ Melvin solution, another  homogeneous magnetic
solution of the Einstein-Maxwell theory (which, however,  is of less interest
since it
does not have $R^4$ topology)
is the Robinson-Bertotti one, i.e. (AdS)$_2 \times S^2$
with  covariantly constant   monopole-type  magnetic
field   $F_{\theta\vp}=b \sin \te $
  on $S^2$. It has an exact  string
 counterpart$^{24}$
which is a product of the two conformal theories:
``(AdS)$_2$"   (SL(2,R)/Z WZW)   and
``monopole"$^{25}$   (SU(2)/Z$_m$ WZW)  ones.
Other monopole-type string  solutions were considered  in$^{26}$.}

It turns out that the above three basic uniform magnetic field  backgrounds
(`constant magnetic field', `$a=1$ Melvin' and
`$a=\sqrt 3 $ Melvin') are exact string solutions to all orders in $\a'$.
The conformal $D=5$ bosonic
$\s$-models which  describe them  are$^{22,14-16}$ are
(the corresponding
superstring and heterotic string models$^{22,14-16}$ have similar structure)
$$ L_{(const)}  =  - \del t \bd t
 + \b \ep_{ij} x^i
\bd  x^j (\del y - \del t)   +  \del x_i \bd x^i   + \del y \bd y +  \del x_3
\bd x_3 + {\cal R} \p_0\
 $$
$$=   - \del t \bd t    +  \b \r^2 \bd \vp  (\del y -\del t)   +  \del \r \bd
\r
+ \r^2 \del \vp \bd \vp +   \del y \bd y +   \del x_3 \bd x_3 +   {\cal R} \p_0
\ , \eq{1.1}
$$
$$L_{(a=1)}     =  - \del t \bd t +  \del \r \bd \r +  {F(\r) \r^2}
(\del \vp  + 2 \a  \del y)  \bd \vp  +
\del y \bd y +
 \del x_3 \bd x_3  +   {\cal R} \p(\r)  \ , \eq{1.2} $$
$$  e^{2(\p -\p_0)} = F(\r) =
(1+ \a^2 \r^2)\inv   \ ,  $$
$$ L_{(a=\sqrt 3)}     =  - \del t \bd t +  \del \r \bd \r +  {\r^2}
(\del \vp  +  q\del y) ( \bd \vp  + q \bd y)  +
\del y \bd y +
 \del x_3 \bd x_3  +   {\cal R} \p_0  \ .  \eq{1.3} $$
Here $x_1 + i x_2 = \r e^{i\vp}, \ \vp \in (0,2\pi)$ are
coordinates of 2-plane orthogonal to the direction of the magnetic field
and $ \ y\in (0,2\pi R) \ $   is the  Kaluza-Klein coordinate
used (the charges of string states are proportional to $R\inv$).
  The the constants
$\a,\b,q$   determine the strength of the  abelian magnetic (and other)
background fields.

The model (1.1) is  a special case of the following model ($u\equiv y -t, \
v\equiv y +t$, $\ i,j= 1,..,D-1$)
$$ L= \del u \bd v + \del x^i \bd x^i  + 2 A_i (x) \bd x^i \del u  +   {\cal R}
\p_0 \ , \ \eq{1.4} $$
where the interaction term  is reminiscent of the open string  coupling.
Indeed, (1.4) is conformal to all orders$^{22}$   if $\del_i F^{ij} =0$,
i.e., in particular, if $A_i = -\ha F_{ij} x^j$ ((1.1) corresponds to
$F_{ij}= \b \ep_{ij}, \ i,j=1,2$).  The conformal invariance of (1.4) is due to
the special
chiral `null' structure of the interaction term.
When $y$ is non-compact (so that  instead of describing a  $D$-dimensional
magnetic background   (1.4)  has $D+1$-dimensional plane wave interpretation)
 and $F_{ij}=const\ $
(1.4),  can be identified with  the Lagrangian
of the WZW model based on non-semisimple algebra
$[e_i,e_j] = F_{ij} e_v, $ $ \ [e_i, e_u]=  F_{ij} e_j,$ $  \ [e_i, e_v]=[e_u,
e_v]=0$ which admits non-degenerate invariant bilinear form,
$(e_i,e_j)=\delta_{ij}, \ $ $ (e_u,e_v)=\ha$
((1.1) corresponds to the $E^c_2$ theory of ref.27).
The solvability of the constant field model  (1.4) or (1.1)
is related to the fact that the path integral
over  $v$ leads to a constraint on $u$ so that  the model  effectively
becomes  gaussian in $x^i$.

Although the models  (1.2) and (1.3) look quite different from (1.1), we shall
 explain below that   all  of them  belong to one 3-parameter ($\a,\b,q)$
class  of   string
models which are conformally invariant and,  moreover,   {\it exactly
solvable}$^{16}$. They can thus be considered as
 closed string analogs
of the  solvable  `open string in constant magnetic field' model.
  In spite of their  apparently non-gaussian  form
they are related  (by  formal duality transformations)  to simpler flat
 models
(this  partially  is the reason for their solvability).
As in the open string case,  here one is able
to express the corresponding  conformal field theory operators in terms
of the free creation/annihilation operators
and to explicitly determine the string spectrum$^{14,16}$.
These models appear to be simpler than   coset CFT's corresponding to
semisimple gauged WZW models (for  reviews of solvable (super)string models
based on  semisimple coset  CFT's  see, e.g., refs.{28,29}).
For example, their
 unitary  is easy to demonstrate because of the existence of a light-cone
gauge.
These models  (together with plane-wave type
 WZW models for non-semisimple groups$^{27,30-34}$)  are thus  among  the
first
few known examples of  solvable unitary
conformal string models with  non-trivial $D=4$  {\it curved  space-time}
interpretation.

Below we shall first  discuss the target space
interpretation of the above models as  representing a class of exact stationary
axisymmetric magnetic flux tube solutions of string effective equations
(Section 2).
Then in Section 3
 we  shall  construct  the  conformal \sms   describing
 the magnetic flux tube  solutions by starting with flat space model  and using
  world-sheet angular
duality.
This will help  to solve the corresponding
classical  string equations
  explicitly,   expressing the string coordinates in terms of free
fields  satisfying `twisted' boundary conditions (Section 4.1).
After  straightforward operator quantization (Section 4.2)  we will   find  the
quantum Virasoro operators.
 It will then be possible to
determine
 the spectrum of states and partition function (Section 5),  in direct  analogy
with how this is done in simpler models like
 closed  string on a torus or an  orbifold,
or   open string in a constant magnetic field. We shall  also discuss
some properties of the spectrum, in particular, the two types of tachyonic
instabilities   present  in this closed string model.
Some concluding remarks (in particular, about  superstring  and heterotic
string generalizations) will be  made in Section 6.
\vglue0.6cm
\leftline{\tenbf 2. Magnetic flux tube solutions of string effective equations}
\vglue0.4cm
We shall be considering   the closed
bosonic  string (or type II superstring)
 theory which has no fundamental gauge fields in a  higher
dimensional space.  The abelian gauge fields  appear upon  toroidal
compactification   when the theory is  `viewed' from four dimensions.
 The conformal $\s$-models   which  describe
 $D=4$  string solutions  with  non-trivial gauge  fields  will thus be
higher dimensional ones.
The simplest case is that  of  $D=5$
 bosonic string \sm action
(with target space fields  not depending on $x^5$)
 which can be interpreted as an  action
of a  $D=4$ string  with an internal degree of freedom (compact Kaluza-Klein
coordinate $x^5$)
which describes the   coupling  to  additional
vector (and scalar)  background fields,
$$ I_5={1\over \pi\alpha '}\int d^2 \s\big[ ( G_{MN} +B_{MN})(X)
\del X^M \bd  X^N  + {\cal R} \p
(X) \big]  \eq{2.1} $$  $$
= {1\over \pi\alpha '}\int d^2 \s\big[  (\hat G_{\m\n} +B_{\m\n})(x)
\del x^\m \bd x^\n + \  e^{2\s(x) } [\del y+ {\cal A}_\m (x) \del x^\m][\bd y+
{\cal A}_\n (x)
\bd x^\n]  $$
$$
 + \  {\cal B}_{\m   } (x) (\del x^\m \bd y- \bd x^\m \del y )  + {\cal R} \p
(x) \big]\ , $$
where $X^M= (x^\m, x^5), \ x^\m=(t, x^i, x^3), \ x^5\equiv y$,
$\ {\cal R} \equiv \four \a'\sqrt \g R^{(2)}$
and
$$\hat G_{\m\n} \equiv  G_{\m\n} - G_{55}{\cal A}_\m {\cal A}_\n
\ , \ \ \ G_{55}\equiv  e^{2\s}\ , \ \ \ \
 {\cal A} _\m\equiv   G^{55}  G_{\m 5}\ ,  \
\ \ {\cal B} _\m \equiv
B_{\m 5}\ .   \eq{2.2} $$
  From the point of view of the
low-energy effective  field theory,
  this decomposition corresponds to starting
 with the $D=5$ bosonic string effective action and
 assuming  that one spatial dimension $x^5$ is compactified on a small
circle. Ignoring the massive Kaluza-Klein modes  one then
finds the following dimensionally reduced  $D=4$  action (see,  e.g., ref.{35})
$$ S_4 = \int d^4 x \sqrt {\hat G }\  e^{-2\Phi}    \ \big[
  \   \hat R \ + 4 (\del_\m \Phi )^2  - (\del_\m \s )^2  \eq{2.3} $$
 $$- \ {\textstyle {1\over 12}} (\hat H_{\m\n\l})^2\  -  \four e^{2\s} ({
F}_{\m\n}
({\cal A}))^2
-\four e^{- 2\s} (F _{\m\n} ({\cal B}))^2
  + O(\a')   \big]  \  , $$
 $$ F_{\m\n} ({\cal A}) = 2\del_{[\m}
{{\cal A} _{\n]}  \ ,   \ \ F_{\m\n}({\cal B}) = 2 \del_{[\m} {\cal B}_{\n]} }
\  ,   \eq{2.4} $$  $$
\hat H_{\l\m\n} = 3\del_{[\l} B_{\m\n]} - 3 {\cal A}_{[\l} F_{\m\n]}
({\cal B})
\   , \ \ \ \ \Phi= \phi - \ha \s \ .    $$
Given a conformal $D=5$ \sm and rewriting  its action as in  (2.1)
 one can  read off
the expressions for the corresponding $D=4$  background
fields which  then  must  represent a solution
of the effective  equations following from   (2.3).
These equations have, in particular,  the following
3-parameter $(\a,\b,q)$ class of stationary axisymmetric
(electro)magnetic flux tube solutions$^{16}$ ($x^\m=(t, \r, \vp, x^3)$)
$$ ds^2_4
=
 -dt^2  +      F(\r) \r^2 (d\vp -\a dt)(d\vp -\b dt)   $$
 $$ -  \ \four F(\r )\tilde F(\r) \r^4  \big[ (\a-\b -2q ) d \vp + q(\a + \b)
dt\big]^2 +  d\r^2 +  dx_3^2 \  ,\eq{2.5}  $$
$${\cal A}  =- \ha  {{\tilde  F}}(\r)  \r^2[ (\a-\b  - 2q ) d \vp + q
(\a+\b)dt] \ , \eq{2.6} $$
$${\cal B} = -\ha  F(\r) {\r^2 } [ (\a + \b ) d \vp - (2\a\b  + q\a - q\b) dt]
\ ,  $$
$$ e^{2(\p-\p_0) } =  F (\r) \ , \   \ \ \  e^{2\s } =   {F(\r) \over
{{\tilde  F}}(\r)}\ , \ \ \ \   B=
  - \ha   (\a-\b) F(\r)  \r^2 d\vp\wedge dt \ , \eq{2.7}
  $$
$$
F (\r) \equiv {1\over 1 +  \a\b {\r^2} } \ , \ \ \
{{\tilde  F}} (\r) \equiv {1\over 1 + q(q +\b-\a)
{\r^2}}\ .  $$
The metric is  stationary  and, in general,  describes a  rotating
`universe'.
For generic values of the parameters the two  abelian  gauge fields
contain  both
magnetic and electric components  with the former being more `fundamental'
(there are no  solutions when both   gauge fields are pure electric).
For simplicity we shall call these solutions `magnetic flux tube backgrounds'.
The three simplest
 uniform  pure magnetic field  solutions  mentioned in Section 1 are
the following special cases (cf.(1.1)--(1.3)): \ \ \ \
 (i) `constant magnetic field' \    ($q=\a=0, \ \b\not=0$):
$$ ds^2_4 = -  ( dt +  \ha \b \r^2 d\vp )^2 + d\r^2 + \r^2 d\vp^2
 + dx_3^2 \ , \eq{2.8} $$  $$
{\cal A} =-{\cal B} =  \ha \b \r^2  d\vp \  , \ \ \ \s=\p-\p_0= 0\ , \   \ \ \
B=\ha \b\r^2 d\vp\wedge dt\ , \ \
\hat H_{tij } = F_{ij} \ ,   $$
(ii) `$a=1$ Melvin' \   ($\a=\b=q\not=0$):
$$  ds^2_4=  - dt^2  + d\r^2 +  F^2(\r)  \r^2 d\vp^2 + dx_3^2 \ ,  \eq{2.9} $$
$$ {\cal A}= -{\cal B} =  \a  F(\r)   \r^2 d\vp\ , \ \ \ B=0\ , \ \ \
 \s=0\ , \ \ \  e^{2(\p-\p_0)}=  F  = (1 + \a^2 \r^2)\inv  \  ,    $$
(iii) `$a=\sqrt 3 $ Melvin' \  ($\a=\b=0, \ q\not=0$):
$$  \ ds^2_4=  - dt^2  + d\r^2 +  \td F(\r)  \r^2 d\vp^2 + dx_3^2 \ ,
\eq{2.10} $$
$$ {\cal A}=  q  \td F(\r)   \r^2 d\vp\ , \ \ \  \ {\cal B}=0\ , \ \ B=0\ , \ \
\
 \p=\p_0   \ , \  \ \ e^{2\s} = {\td F}\inv = 1 +  q^2 \r^2 \  .    $$
In addition to the $q=0$ subclass of pure
magnetic backgrounds (where ${\cal A}$ has constant field strength) which
generalize (2.8),
there are two other special subclasses:
$\a=q$ (stationary metric, non-zero $B_{\m\n}$, zero $\s$)
and $\a=\b$ (static metric, zero $B_{\m\n}$, non-zero $\s$).
Solutions with $\a\b \geq 0, \ q(q +\b-\a)\geq 0$ have no curvature
singularities.

The above leading-order solutions  (2.5)--(2.7)  are actually
exact  to all orders in $\a'$ since it turns out that they correspond
   (according to (2.1))  to  conformal $D=5$ $\s$-models  discussed in the next
section.\footnote{$^b  $}{\baselineskip6pt
This class of  solutions was actually  found$^{16}$   not by solving the
complicated equations which follow from
(2.3) but   by explicitly
constructing  the corresponding $D=5$ conformal \sm discussed below.
Similar approach to constructing exact string solutions
was  used in ref.{22}.}
\vglue0.6cm
\leftline{\tenbf 3. Conformal string models describing   flux tube backgrounds
}
\vglue0.4cm
 The three conformal  $D=5$ \sms  that correspond  to the $D=4$
solutions (2.8),(2.9),(2.10)  according to (2.1)  are  indeed
(1.1),(1.2),(1.3).
All three models have free-theory central charge.
In the case of non-compact $y$, i.e. in the limit $R\to \infty$,  they
are equivalent to  other known models.
The constant field model (1.1)  becomes the `plane-wave' $E^c_2$ WZW
model$^{27}$  with
the corresponding CFT discussed in refs.31,14,36.
The $a=1$ Melvin model$^{15}$  (1.2)
 with  coordinates  formally taken to be non-compact  can be identified with a
particular   limit
of  $[SL(2,R)\times R]/R$ gauged WZW  (`black string'$^{37}$) model\footnote{$
^c$}{\baselineskip6pt
In this limit $k\to \infty$ and the mass and charge
of `black string' vanish but simultaneous rescalings of coordinates
give rise to a nontrivial model.}
 or,  equivalently,  with the    $E^c_2/U(1)$ coset  theory$^{32}$.
 The $R= \infty$ case of the $a=\sqrt 3 $ Melvin model (1.3)
is identical  to the flat space model after the redefinition of $\vp$.

The solvability of the
 general 3-parameter $(\a,\b,q)$ class of string models corresponding
to   (2.5)--(2.7)
can be  understood  by  using
 their relation via duality and formal coordinate shifts to  flat space
models.
Consider, for example,  the \sm
which a direct product of $D=2$ Minkowski space and $D=2$ `dual 2-plane'
$$  \tilde I={1\over \pi\alpha '}\int d^2 \s\big[ \del u \bd v
+ \del \r \bd \r + \r^{-2} \del \tvp\bd  \tvp  + {\cal R} (\p_0 - \ln \r)\big]
\  .  \eq{3.1} $$
 $\tvp$  should have
period $2\pi\a'$ to preserve equivalence of the `dual 2-plane' model to the
flat 2-plane CFT$^{38}$, i.e. to the  flat space  model\footnote{$^d
$}{\baselineskip6pt
The two models are equivalent in the sense of a relation of classical solutions
and equality of the correlators  of certain operators  (e.g., $\del \tvp$ and
$\r^2 \del \vp$)
 but the spectra of states  are formally different (see  also ref.{39}): the
spectrum is continuous on 2-plane and discrete on dual 2-plane (with  duality
relating
 states with given orbital momentum on 2-plane   and  states with given winding
number on dual 2-plane).}
 $$ I_0={1\over \pi\alpha '}\int d^2 \s\big(  \del u \bd v
+ \del \r \bd \r + \r^{2} \del \hat \vp\bd \hat  \vp  + {\cal R} \p_0  \big)\ .
\eq{3.2} $$
If we now  make coordinate shifts  and add a  constant antisymmetric tensor
term we obtain from (3.1) ($\a,\b, q$  are  free parameters of dimension
$cm^{-1}$)
 $$ \td I={1\over \pi\alpha '}\int d^2 \s \big[
( \del u + \a \del \tvp)  (\bd v + \beta  \bd \tvp)
+ \del \r \bd \r + \r^{-2} \del \tilde \vp \bd \tilde \vp
$$  $$ + \  \ha q [\del ( u  +v) \bd \tvp - \bd (u +v) \del \tvp]
+ {\cal R} (\p_0 - \ln \r ) \big] \ . \eq{3.3}   $$
The  two models  (3.1)  and (3.3)   are of
course `locally-equivalent'; in particular,
(3.3)  also  solves the  conformal invariance equations. However,
if $u$ and $v$  are  periodic, i.e. if $u$ and $v$ are given by
$$u\equiv  y-t\ , \ \ \  \  v\equiv y + t \   , \ \ \  \
y\in (0,2\pi R) \ ,  \eq{3.4} $$
then
the
`shifted' coordinates $u + \a \td \vp $ and $v + \b \td \vp$
are not   globally defined  for generic $\a$ and $\b$  (the periods of $y=\ha
(u + v)$ and $\vp$ are different) and the torsion term is non-trivial
for $q\not=0$. As a result,
the conformal field  theories corresponding to
(3.1) and  (3.3)   will  {\it not}  be equivalent.
The $O(3,3;R)$ duality transformation  with  {\it continuous }  coefficients
which relates the model (3.3) to the  flat space
one (3.2)  is not a symmetry of the flat CFT, i.e. leads to a new conformal
model
which, however, is simple enough to be  explicitly solvable$^{16}$.

Starting with  (3.3)   and making the
duality transformation  in $\tvp$
one obtains a more complicated \sm\  ($\pi \a' I \equiv  \int d^2\s L $)
$$  L=
 F(\r) (\del u  -\a  \r^2 \del \vp') (\bd v   +  \b \r^2 \bd \vp'  )
+ \del \r \bd \r +
   \r^2 \del \vp'  \bd \vp'   \eq{3.5} $$
$$  + \  \del x_3 \bd x_3 +   {\cal R} (\p_0 +  \ha \ln F )\  ,    $$
$$
 F\inv = 1 + \a\b \r^2\ , \ \ \ \  \ \  \vp'\equiv \vp + \ha q(u+v) \ .  $$
Here $\vp$  $ \in (0,2\pi) $ is  the periodic
coordinate dual to $\tvp$ and we have also added a free $x_3$-coordinate term.
Since the periods of $\vp$ and $y= \ha (u+v)$ are, in general,
different $\vp'$ is not   globally defined.
The theory  (3.5)   is conformally invariant to all orders in $\a'$.
For the purpose of  demonstrating  this  one may ignore the difference between
$\vp'$ and $\vp$ (i.e.  may set $q=0$ or consider $y$ to be non-compact).  Then
(3.5)   becomes equivalent to  a special case of the `generalized $F$-model'
which  was  shown to be conformally invariant$^{22}$.

It is the \sm
(3.5) that  defines the string theory    corresponding  to  the  class of
$D=4$ magnetic flux tube  backgrounds  (2.5)--(2.7).
 The models  (1.1),(1.2),(1.3)
are  the    special cases of (3.5): ${\a=q=0}$, $\  {\a=\b=q}\
$ and $ {\a=\b=0}$).
\vglue0.6cm
\leftline{\tenbf 4. Solution of the  string models  }
\vglue0.4cm
\def \X {{\cal X}} \def \t {\tau}
The relation  of  the  class of  string models (3.5)
to the flat space model  via formal duality and coordinate shifts
makes possible  to solve the classical  string equations
(which  look quite complicated)
   explicitly. The fact that  the two dual models have related classical
solutions
enables to
  express the solution in terms of free
fields  satisfying `twisted' boundary conditions$^{14,16}$.
One can then   proceed to
 straightforward operator quantization (fixing, e.g.,  a `light-cone'  gauge).
Some of the resulting expressions are  similar to  those
 appearing in the simpler cases of open string theory in a constant magnetic
field$^{8}$
 or $R^2/Z_N$ orbifold model$^{40}$.
\vglue0.3cm
\leftline{\tenit 4.1.  Solution of the classical equations on the  cylinder}
\vglue0.3cm
Introducing the free field  $X=X_1 + iX_2$
such that
$$ L_0=\del_+\rho\del_-\rho+ \rho^2 \del_+\hat \vp \del_-\hat \vp =
\del_+ X \del_- X^* \ ,
\ \ \  \ \ \  X\equiv  \r e^{i\hat \vp} \ ,   $$
$$ \r^2 = XX^*\ , \ \ \ \hat \vp = {1\over 2i}\ln{X\over  X^*}\ , \ \ \
 X=X_+  (\s_+) + X_- (\s_-) \ , \ \ \s_\pm = \tau \pm \s \ ,   \eq{4.1} $$
we can represent  the solution of equations following from (3.1) in the form
$$ \del_\pm  \tilde  \vp= \mp \r^2 \del_\pm  \hat \vp=\pm {i\over
2}(X^*\del_\pm
X-X\del_\pm X ^*)\  , $$
$$
\tilde\vp(\s,\tau  )= 2\pi\a' [J_-(\s_-) - J_+(\s_+)]  +{i\ov 2} \big( X_+
X^*_-
-X_+^*  X_-\big)\ , \eq{4.2} $$
$$
J_\pm(\s_\pm)\equiv {i\ov 4\pi\a '}\int_0 ^{\s_\pm}d\s_\pm\big( X_\pm \del_\pm
X^*_\pm  -
X_\pm^* \del_\pm  X_\pm\big)\ .  $$
Then  the solution of the string equations corresponding to (3.5)  is
$$
u=U_+ +U_- -\a\tilde\varphi \ ,\   \ \ v=V_+ +V_- -\b \tilde\varphi\  ,
\ \ \  x \equiv \r e^{i\vp} = e^{-iq (u + v )} e^{i\a V_- -
i\b U_+    }  X  ,  \eq{4.3}  $$
where $U_\pm $ and $  V_\pm $ are arbitrary functions of $\s _\pm $.
The closed string boundary condition on the cylinder
$\ x(\s + \pi, \tau)=x(\s, \tau)$ implies that the free field $X=X_+ +
X_-$ must satisfy  the   ``twisted"    condition
$$ X(\s + \pi, \tau)= e^{ i\g \pi } X(\s,\tau)\ , \ \ \  \
 X_\pm  = e^{\pm i\g \s_\pm } \X_\pm  \ ,  \ \ \  \X_\pm (\s_\pm \pm \pi)=
\X_\pm (\s_\pm )\ ,  \eq{4.4} $$
where $\X_\pm = \X_\pm  (\s_\pm)$ are free  fields with standard periodic
boundary conditions
$$ \X_+  =  i  \sqrt{\a'/ 2 } \sum_{n=-\infty}^\infty \tilde a_n
\exp (-2in \s_+)
  \  , \ \ \ \
\X_-  =  i  \sqrt{\a'/ 2 } \sum_{n=-\infty}^\infty  a_n
\exp (-2in \s_-)
  \ . \eq{4.5} $$
Since  $y=\ha (u+v) $ is compactified on a circle of radius
$R$,
$$
u(\s+\pi, \tau)=u(\s,\tau)+2\pi wR\ ,\   \ \ v(\s+\pi, \tau)=v(\s,\tau)+2\pi w
R\
, \ \ \ w=0, \pm 1, ...,  $$
where $w$ is  the  winding number.
As a result,
$$
U_\pm =\s_\pm p_\pm ^u    + U_\pm '\ ,
\ \ \ \ V_\pm =\s_\pm p_\pm^v    + V_\pm'\ , \eq{4.6}
$$
$$
p_\pm^u= \pm ( wR - \a' {\a}J)+p_u  \ ,\ \ \
p_\pm^v= \pm(wR -  \a' {\b  }J)+p_v  \  ,
$$
where $U_\pm' $ and $V_\pm' $ are single-valued functions of $\s_\pm$, \
$p_u$ and $p_v$   are arbitrary  parameters (related to the   Kaluza-Klein
momentum and the  energy of the string) and $J$ is the  angular momentum
($J_{L,R}\equiv J_\pm (\pi )$)
$$  J=J_R+J_L =  -\ha\sum _n (n+\ha\gamma ) a^*_na_n -\ha \sum_n (n-\ha\gamma )
\td a^*_n \td a_n     \ . \eq{4.7}  $$
Then the   `twist' parameter  $\g$ in (4.4)  is given by
$$ \gamma =(2q + \b-\a )wR+\b p_u+\a p_v
\ .  \eq{4.8} $$
Evaluating   the classical
stress-energy tensor   on the above  solution
one finds that it takes the ``free-theory" form
$
T_{\pm\pm}= \del_\pm U_\pm \del_\pm V_\pm   + \del_\pm
X \del_\pm X^*.$
 It is convenient to fix the light-cone gauge, using the remaining  conformal
symmetry
to gauge away  the `non  zero-mode' parts   $U'_\pm $ of $U$.
Then the  classical constraints $T_{--}=T_{++}=0$ can be solved
as
usual and  determine $V'_\pm $
  in terms of the free fields  $X_\pm$.
The classical expressions for  Virasoro operators
 $L_0= {1\over 4\pi \a'} \int d\s T_{--},$  \ \ $  \tilde L_0= {1\over 4\pi
\a'}\int d\s T_{++}$
 are
$$L_0=   {p_-^up_- ^v \ov 4\a' }
 +\ha\sum_{n }  \big(n+\ha \gamma \big) ^2 a_n^{*} a_{n}\  , \ \ \
\tilde L_0 =    {p_+^u p_+^v\ov
4\a'} + \ha
\sum_{n}   \big( n-\ha  \gamma  \big) ^2      \td a_n^{*} \td a_{n} \ .
\eq{4.9} $$
\vglue0.4cm
\leftline{\tenit 4.2.  Quantum Virasoro  operators }
\vglue0.4cm
One can now
 quantize the theory by imposing  the  standard  commutation relations
between canonical momenta and coordinates $ [P_x(\s,\t), x^*(\s',\t)]=
-i\delta(\s -\s')$, etc. Because of the duality between (3.3) and (3.5)
 this  turns out to be the same as
 demanding the canonical commutation relations
for the  fields $X, X^*$ of the free (but globally non-trivial,
cf.(4.4)) theory.
As a result,  $p_u, p_v$  and the Fourier modes  $a_n, \td a_n$
 become  operators  in a  Hilbert  space. One finds that
$$ [ a_n, a_m^{*}] =    2  (n+ \ha \gamma  )\inv   \delta _{nm}\ , \ \ \  \ \
[ \td a_n, \td a_m^{*}] =    2  (n - \ha \gamma )\inv   \delta _{nm } \ .
\eq{4.10} $$
 $p_u, p_v $   and thus $\g$ (4.8)
commute with $a_n,\ \td a_n$  and   can
 be expressed in terms of the
conserved  string energy $ E=  \int_0^\pi d\s P_t$ and  the  quantized
Kaluza-Klein linear momentum $ p_y= \int_0^\pi d\s P_y = { m/ R}, $
\  $m=0, \pm 1 , ..., $
$$
E=   { {1\ov 2\a '}} [p_u - p_v  -   \a'(\a + \b ) {\hat J}]   \ ,\ \
\   p_y=  { {1\ov 2\a '}} [ p_u + p_v + \a'   (2q + \b - \a ) \hat J]   \ ,   \
\eq{4.11} $$
$$
\gamma =(2q + \b -\a)wR +\a'[(\a + \b )m R \inv - (\a -\b )E] -  \ha \a 'q(\a +
\b)\hat J\ . \eq{4.12}
$$
  $\hat J$ is the angular momentum operator  obtained by
`symmetrizing' the classical $J$ (4.7).
The quantum Virasoro operators $\hat L_0$ and ${\hat {{\td L}}}_0$
(and thus the quantum Hamiltonian $  \hat H  = \hat L_0 + {\hat {{\td L}}}_0 $)
are then given by symmetrized  expressions  in  (4.9).\footnote{$^e
$}{\baselineskip6pt In agreement with  the defining relations in (4.4)   the
expressions for $\hat H$, $\hat J$ and the commutation relations  (4.10)
are invariant under $\g \to \g +2$ combined with the corresponding renaming
of the mode operators $a_n \to a_{n+1}, \ \td a_n \to \td a_{n-1}$. }

The  sectors of states    can be  labeled  by  the conserved
quantum numbers:  the   energy $E$, the angular momentum  $\hat J$ in the
$x_1,x_2$ plane,  and the
linear $m/R$ and winding  $wR$  Kaluza-Klein  momenta  or  ``charges"
(and also  by momenta in additional 22 spatial dimensions which we shall  add
to
saturate the central charge condition).
As in the case of the Landau model or the open string model$^{8}$, the states
with  generic values of
$\g$ are ``trapped" by the magnetic field.
 The  states in  the
 ``hyperplanes" in the $(m,w,E,\hat J)$ space
with  $|\gamma| = 2n$, $ n=0,1,...,$
are special:
 for them  the  translational invariance on the $(x_1,x_2)$-plane is restored
with the   zero-mode  oscillators $a_0, a_0^*,  \td a_0, \td a_0^*$  being
replaced by the zero mode part of the coordinate  $x$ and conjugate  linear
momentum.

Restricting the consideration to the sector of states with
 $ 0 <  \g  <2 $
 one can   introduce  the  normalized creation and annihilation
operators ($n,m=1,2,...$)
$$
[ b_{n\pm}, b_{m\pm }^{\dagger}] = \delta _{n m}  ,\ \
 [\td b_{n\pm}, \td b_{m\pm }^{\dagger}] = \delta _{n m}  ,\ \
[b_0,b_0^{\dagger}]=1 \ , \ \ [\td b_0,{\td b}_0^{\dagger}]=1\   ,  \eq{4.13}
$$
$$ b_{n+}^{\dagger}= a_{-n} \omega_-  ,   \   b_{n-}= a_{n} \omega_+ ,   \
b_0=\ha \sqrt{\gamma }a_0
,   \ {\td b}_0^{\dagger}=\ha \sqrt{\gamma } \td a_0 , \
\omega_\pm \equiv \sqrt {  \ha \big( n \pm \ha {\gamma } \big) }. $$
 The Hamiltonian and Virasoro operators  then  take the form$^{16}$
$$
\hat H= {\hat L}_0 + {\hat {{\td L}}}_0=
 \ha  \a' \big( -E^2 +  p_a^2 + \ha Q_+^2 +
\ha Q_-^2  \big) + N+  {\td N}-2c_0
\  \eq{4.14} $$
$$ -
\a'   [(q +\b)Q_+  + \b  E] J_R - \a'[ (q-\a) Q_-  + \a E] { J}_L $$
$$  + \
 \ha \a'q\big [  (q+2\b)  J_R^2 + (q-2\a ) J_L^2
+  2 (q + \b -\a )  J_R  J_L  \big] \ ,   $$
$$ {\hat L}_0 - {\hat {{\td L}}}_0 = N-\td N -mw \ .  \eq{4.15} $$
Here $Q_\pm$ are the left and right combinations of the Kaluza-Klein  linear
and winding momenta (which play the role of charges in the present context),
$c_0$ is the normal ordering term\footnote{$^f  $}{\baselineskip6pt The  normal
ordering constant
is  fixed by  the Virasoro algebra. The free-string
constant in  $\hat L_0$ is shifted  from 1 to
$c_0 $.
This corresponds to computing   infinite sums  using   the generalized
$\zeta$-function
regularization.
 Similar shift  is found in the
 open
string theory$^8$   and in orbifold models and  is
characteristic  to the case of a free boson  with twisted boundary conditions.
It is also  consistent  with  modular
invariance  of the partition function. }
$$ Q_\pm \equiv {m\ov R} \pm  {wR\ov \a'} \  , \ \  \ \
   \   \
 \ \ c_0 \equiv 1- \four \gamma(1-\ha \gamma) ,
$$
and $p_a, \ a=3, ..., 24\ $ are   momenta  in   additional  free spatial
dimensions.
The operators $N,\td N$  and  the angular momentum  operators  $J_L,J_R$
have the  standard `free-theory' form
($a_{na}, \td a_{na}$  correspond to free spatial directions, $a=3,...,24$)
$$
 N= \sum_{n=1}^\infty n ( b^{\dagger}_{n+}b_{n+}+ b^{\dagger}_{n-}b_{n-}
+ a^{\dagger}_{na} a_{na} ) \ , \ \
 {\td N}= \sum_{n=1}^\infty n ( \td b^{\dagger}_{n+}\td b_{n+}+ \td b^{\dagger
}_{n-}\td b_{n-}+\td a^{\dagger}_{na} \td a_{na}   ) , $$
$$ {\hat J}_R= - b^{\dagger}_0 b_0  - \ha  +\sum_{n=1}^\infty \big(  b^{\dagger
}_{n+}b_{n+} - b^{\dagger}_{n-}  b_{n-} \big)\equiv J_R-\ha \to -l_R - \ha +
S_R
 \ , \eq{4.16} $$
$$
{\hat J}_L= \tilde b^{\dagger}_0 \tilde b_0  + \ha  +\sum_{n=1}^\infty \big(
\tilde b^{\dagger
}_{n+}\tilde b_{n+} - \tilde b^{\dagger}_{n-} \tilde b_{n-} \big)\equiv J_L +
\ha
  \to l_L  +  \ha + S_L
  , \ \
\hat J =J_R + J_L =J . $$
The  analogs of the  above expressions
  in the sectors with $2k<\gamma <2k +2,\ k=$integer,  can be found in a
similar way   by  `renaming' the creation and annihilation operators.
The result is the same as  in  (4.14)  with the replacement
$ \g \to \g '= \g-2k $ in $c_0$.

It is remarkable that the complicated space-time background (2.5)--(2.7)
 is associated with relatively simple CFT described  by
(4.14).
The first line in (4.14)  with $c_0 \to 1$ is the Hamiltonian
of a  free  closed string compactified on a circle.
The second line (together with $O(J)$ term in  $c_0$, see (4.12))  is the
analogue of the gyromagnetic interaction term
for a particle in a magnetic field.\footnote{$^ g $}{ \baselineskip6pt
The $EJ_{L,R}$  and $\a'E^2$ terms (explicit in (4.14)
 and implicit in $c_0$ through its dependence on $\g$) reflect
the non-static nature of corresponding subclass of  backgrounds (which in turn
is
related  to the presence of the non-vanishing antisymmetric tensor).}
  Similar term is present in the   Hamiltonian of the open string  in a
constant magnetic field$^{8}$
$$
\hat H^{(open)} =L_0=
 \ha  \a' \big( -E^2 +  p_a^2) + N - c_0 (\g)   - \g J_R
\ , \eq{4.17}
$$
$$ c_0 =1-\four \g  (1- \ha\g ) \ , \ \ \   \g \equiv {2 \ov \pi} |{ arctan}
(2\a' \pi Q_1 \b) +
{arctan} (2\a' \pi Q_2  \b )| \ , $$
where $Q_1,Q_2$ are charges at the two ends of the open string, $N$ and $\hat
J_R$ have the same form as in (4.16)
and $\b$ is proportional to
 the magnetic field, $F_{ij}= \b \epsilon_{ij}$.
The $O(J^2)$ terms in the  third line of (4.14)   (and in $c_0$)
are special to closed string theory.\footnote{$^h  $}{ \baselineskip6pt It is
clear from the above construction that (4.14) is, at the same time, also
the Hamiltonian for the  $\vp$-dual theory (3.3). The  origin of the $J^2$
terms in $\hat H$ can be  traced, in particular,
 to the presence of the $\a\b \del \td \vp \bd \td \vp $ term in (3.3).}

In the special cases corresponding to the
`constant magnetic field' model (1.1), (2.8),
and  the $a=1$  and $a=\sqrt 3 $ Melvin models (1.2), (2.9)   and (1.3), (2.10)
the Hamiltonian (4.14) takes the following form
$$\a=q=0 \ : \ \ \
\hat H = \hat H_0 -2c_0 (\g)  -
\a' \b   (Q_+  +  E) J_R  \ , \ \ \ \g=\a'\b (Q_+ + E) \ , \eq{4.18}  $$
$$  \hat H_0 \equiv \ha   \a' \big( -E^2 +  p_a^2 + \ha Q_+^2 +
\ha Q_-^2  \big) + N+  {\td N}\ ,
$$
$$
{\a=\b=q}\ : \ \ \ \hat H=\hat H_0
 -2c_0(\g)  -
2\a' \a   Q_+   J_R - \a'\a  E ( J_R + J_L)  \ \ \ \ \ \ \ \ \ \ \ \ \
  \eq{4.19} $$
$$ \ \ \ \ \ \ \ \ \ \ \ \ \ \ \ \ \ \ \ \ \  + \
 \ha \a'\a^2  ( J_R + J_L)(3J_R - J_L )   \ ,   \ \ \ \ \  \g = \a'\a [ Q_+  -
\a( J_R + J_L )]  \  , $$
$$
{\a=\b=0}\ : \ \ \  \ \hat H = \hat H_0
-2c_0(\g) -
\a' q (  Q_+  J_R  + Q_-  { J}_L)  +
 \ha \a'q^2  (J_R + J_L) ^2 \ , \eq{4.20}   $$
$$  \ \ \  \   \g= 2q wR \  . $$
Note that   presence of the $O(\g^2)$ normal ordering term in $c_0$ in (4.14)
implies (see (4.12)) that the quantum Hamiltonians (4.14),(4.18)--(4.20)
contain  just one  $O(\a'^2)$ term
which is of    higher order in $\a'$ than  other `semiclassical'
terms.\footnote{$ ^i $}{ \baselineskip6pt The presence of this  higher order
term is consistent with
current algebra  approaches in the  two special cases when  our   model
becomes equivalent to  a
WZW or coset model:
 (i) $R=\infty$ limit of the
constant magnetic field model (1.1) which is   equivalent  to
the $E^c_2$ WZW model$^{27,31,32}$  (for which the quantum stress tensor
contains  order $1/k  $ correction  equivalent to the term  in (4.18) in this
limit);
\ \  (ii) the non-compact limit of the $a=1$
Melvin model (1.2)   which is  related   to  a  special limit of the
$SL(2,R)\times R/R$
gauged WZW model, or  to the    $E^c_2/U(1)$ coset  theory,  the
 Hamiltonian of which also contains  $1/k$ correction term$^{32}$.}
\vglue0.6cm
\leftline{\tenbf 5.  String spectrum and partition function }
\vglue0.4cm
Using (4.14),(4.15) to define the Virasoro constraints
$$ \hat L_0=
{\hat {{\td L}}}_0 =0\  \ \ \   \to   \ \ \ \  \hat H=0\   ,  \
\ \ \ \  N- {\td N}=mw  \ ,  \eq{5.1} $$
it is straightforward to compute the spectrum of  states$^{14,16}$
just as this is done in  the free string theory.
Indeed,  even though the
Hamiltonian (4.14) containing $O(J^2)$ terms
   is, in general,    of  {\it fourth}
order in creation and annihilation operators,
it
is  {\it diagonal}  in Fock space   since  $N, \td N, J_L$ and
$J_R$ have  diagonal form.
The continuous
momenta $p_{1,2}$  corresponding to the zero modes of the
coordinates $x_{1,2}$  of the plane are  effectively replaced  by the integer
eigenvalues
$l_R, l_L =0,1,2,...$ of the zero-mode parts
$b_0^{\dagger}b_0$ and $\td
b_0^{\dagger}\td b_0$   of $\hat J_R$ and
$\hat J_L$ (see (4.16)). Thus  the `2-plane'  part of  the spectrum is
discrete
in the $0<  \g < 2  $   sector
(but, as mentioned above, becomes   continuous  when  $\g=0$ or
$\g=2$).\footnote{$ ^j $}{\baselineskip6pt
 The Hamiltonian for the case of $\g=0$ is obtained by adding
$\ha\a'(p_1^2+p_2^2)$ and replacing
$-b_0^{\dagger}b_0-\ha $ and $\td
b_0^{\dagger}\td b_0 + \ha $   in  $\hat J_R$ and
$\hat J_L$ in (4.16)  by one half  of the  center of mass orbital momentum
$(x_1p_2-x_2p_1)$.}
 Generic string states  are thus  `trapped'  by the magnetic  flux tube
 as in  the   Landau problem  or  the open string case.
This  is  consistent with   a picture of a charged  closed string moving
 in   magnetic field
orthogonal to the plane.

For example, let us consider the
scalar state at zero string excitation
level $S_L=S_R=N=\td N=0$ in the non-winding ($w=0$) sector.
The eigen-values of $\hat J_R$ and $\hat J_L$ in (4.16)
are $-l_R -\ha $ and $l_L + \ha$  ($l_{L,R}=0,1,2,...$ are the analogs of the
Landau level).
Then  in the $a=1$ Melvin model (4.19)
$\hat H =0$ reduces to
$$ M^2\equiv E^2- p_a^2 = - 4{\a'}\inv    + p_y^2  + 2\a p_y (2l_R + 1)
- 2\a^2 (l_L-l_R)(2l_R + 1)  \eq{5.2} $$
$$  - 2\a' \a^2[p_y - \b (l_L-l_R)]^2
 \ , \ \ \ \ \ \ \    p_y= m/R\ , $$
where it is assumed that
 $ 0< \g=  2\a' \a  [p_y - \a (l_L-l_R)] <2. $
The same   spectrum (up to the $O(\a')$ term   coming  from
$\g^2$   in $c_0$ in (4.14)) can be found  by
  directly solving  the  tachyon equation
(to the leading order in $\a'$)
 $$ \a' [\Delta + O(\a')]  T = 4 T  \ , \ \ \ \ \   \Delta = - {1\ov{\sqrt{ -G}
e^{-2\p} }} \del_\m ({\sqrt{ -G}  e^{-2\p} }G^{\m\n} \del_\n) \ .
  $$
In  the $D=5$ background corresponding to the  Melvin model (1.2),
$$ \big[ - \del_t^2  + \r\inv \del_\r(\r\del_\r)
+ \r^{-2}(1 + \a^2\r^2)^2
\del_\vp^2
+    (1 + \a^2\r^2) \del_y ( \del_y
-2\a \del_\vp) \big] T=
-4{\a'}\inv  T  \ ,  \eq{5.3} $$
and (5.2) is reproduced by taking
 $T= \exp (iEt + ip_y y + il\vp ) \ \tilde T(\r)  , \ l = l_L-l_R.$
Similar correspondence between the string spectrum and the solution of the
tachyon equation
 is found  also in the constant magnetic field model (1.1)
where the point-particle limit of the Hamiltonian (4.18)
  is
$$  \hat H_0 = {\ha }\a'   \big[- E^2+ p_a^2 +
{ p}_y^2     +2 ( { p}_y +  E)\b (l_R + \ha)
-\ha \a' \b^2(p_y +E)^2 \big] -2       \ .  \eq{5.4} $$
In semiclassical approximation  (5.4)
 is similar to the Landau Hamiltonian  with ${ p}_y +  E$ playing the role of
charge.  The unusual  dependence on the energy
is due to the fact that the background (2.8)
is stationary but not static (or, equivalently, due to the presence of the
antisymmetric tensor background as  demanded by conformal invariance).
For the subclass of  models
with $\a=\b$ the metric is static (and $B_{\m\n}=0$,  see (2.5),(2.7))
and we get  more direct correspondence  with familiar particle
theory  expressions.
Indeed, in the weak-field limit when $\a=\b$ and $q$ are small
one finds from (4.14) (cf. (5.2))
$$
 M^2= M^2_0-2(q_+Q_+S_R+q_-Q_-S_L) +
[(2l_R+1)q_+Q_+-(2l_L+1)q_-Q_-]+O(q_\pm^2)
\ ,  $$
$$
\a'M_0^2=-4+2N+2\td N+\ha Q_+^2+\ha Q_-^2 \ , \ \ \ \ \ \ q_\pm = q \pm \a \ .
$$
 Since in the {\it closed} string  models we consider there are
 {\it two}  $U(1)$ gauge fields (2.6)
with
strengths determined by $q_\pm$ we  get two
gyromagnetic ratios (in general, different from 2),
  $g_R= 2S_R/S, \ g_L= 2S_L/S, \  \  S= S_R + S_L$,
which are in agreement$^{16}$  with  earlier suggestions ref.{13}.
\vglue0.6cm
\leftline{\tenit 5.1. Tachyonic instabilities }
\vglue0.4cm
The effect that the magnetic field produces on the energy of a generic
state  can be interpreted as  a combination of the gyromagnetic Landau-type
interaction and the influence of  curved
space-time geometry.
The   important  property  of the spectrum is   the appearance of
new tachyonic instabilities, typically associated with states with
angular momentum aligned along  the magnetic field.
Similar  magnetic instabilities were found  in  non-abelian  field
theories$^{5,6}$ (where one has charged spin 1 particles with non-minimal
coupling) and in  the  (abelian)
open string theory$^{8}$.  In the open string case there  is a sequence of
critical
values of the magnetic field  for which highest spin component states
(lying on the first Regge trajectory, $(b^{\dagger
}_{1+})^k|0; l_R=0 >$, cf. (4.16),(4.17)) become tachyonic$^{10}$.
The  new feature of
 the   closed string theory  is the existence
    of states
 with arbitrarily large charges.
Since the critical
magnetic field at which a given state of a charge $Q$  may become tachyonic
is of order of $1/(\a' Q) $ there is  an infinite number  of  tachyonic
instabilities
for any given finite value of the magnetic field.
Also,  in contrast to the abelian
open (super)string model$^{8,10}$
where  charged $S \geq 1$  spin  states are
massive and thus instability  can appear only
for  large ($\sim 1/\a'$)  values of the magnetic field, the
free closed string spectrum contains charged massless states
so that (as in the unbroken gauge theory$^5$)
 tachyonic states appear   for an infinitesimal value of the  background
magnetic field.

To illustrate the presence of the new tachyonic instabilities
 let us  consider the constant magnetic field  model (4.18)  and look at   the
states which
complete the $SU(2)_{R}$ massless vector
multiplet in the free ($\b =0$)  theory
compactified on a circle of `self-dual' radius $ R=\sqrt{\a'}$. The states with
$S_R\neq 0$ are
$
b^{\dag }_{1\pm }|0;m=w=1\rangle  ,\ \ b^{\dag }_{1\pm }|0;m=w=-1\rangle , $
i.e. have $\td N=0$,  $\ J_R=-l_R \pm 1  $,
and the  energy
$$
  \kappa\big[ E + {\kappa\inv \b  }({\hat J}_R+\ha \a'\b  Q_+ )\big]^2 =
-4{\a'}\inv   +
{\kappa\inv } (Q_+ - \b  {\hat J}_R)^2  \ ,    \eq{5.5} $$
$$ \kappa \equiv  1 + \ha \a' \b^2 \ , \ \   \ \ \ Q_+ =
(r+ r\inv )/\sqrt{\a'} \ , \ \  \   \ \ r\equiv R/\sqrt{\a'}  \ . $$
If $r=1$, an infinitesimal magnetic field
 $\b >0$  makes
the component with $J_R=1$   tachyonic. This instability is the same as
in the
 non-abelian gauge  theory$^5$.
Away from the self-dual
radius, this state  is mahas real  energy for small $\b$ and becomes tachyonic
for
 some  finite critical value $\b_{cr}$ of the magnetic field.

 Instabilities caused by the linear in $\hat J_{L,R}$ terms
in $\hat H$ are present also in the $\a=\b$ models, in particular, in the
$a=1$ Melvin model (4.19).
For example, the mass of the  level one   state
with $w=0$, $m >0 $, $\  N=\td N=1$,  $l_R=l_L=0, \  S_R=1, S_L=- 1$
(which  corresponds to  a   `massless' scalar with  a  Kaluza-Klein charge) is
 (cf. (5.2))
$ M^2=  p_y(p_y -2 \a -2\a'\a^2 p_y )$.
For large enough $R$, \ $M^2$ becomes negative when $\a >\a_{ cr}\sim \ha p_y$.
For these states   $\g =2\a' \a p_y  $ and thus $\g <2$ if
 $\a >\a_{cr}$ and  $\a' p^2_y <2$.
The critical value of the magnetic field goes to zero as $R\to\infty
$.\footnote{$^k  $}{\baselineskip6pt In the noncompact  case   $p_y$ becomes a
continuous parameter representing the momentum of the  `massless' state in the
$y$-direction.
Thus the `massless' state
 with  an infinitesimal momentum $p_y$ becomes tachyonic for an infinitesimal
value of $\a $.}
The  $a=\sqrt {3}$  Melvin  model is stable in the non-winding ($w=0$, i.e.
$\g=0$) sector, in the sense that  it has  no new  instabilities in addition to
the usual flat space tachyon.  For $w\not=0$ (and $0< \g <2$)
there exists a range of parameters $q,\ R\  $
for which  there is again the same linear instability as
 in the $a=1$ Melvin model$^{16}$.

\def \h {\chi}
\def \td {\tilde}
\vglue0.6cm
\leftline{\tenit 5.2.  String partition function on the torus}
\vglue0.4cm
Given the explicit expressions for the Virasoro operators in
(4.14),(4.15)
it is straightforward to
compute the partition   function of this conformal model,
$$Z=\int {d^2\t\ov \t_2} \int dE\prod_{a=3}^{24} dp_a
\sum_{m,w=-\infty}^\infty  {\rm Tr}  \exp \big[ 2\pi i( \t {\hat L_0} - \bar \t
{\hat
{\td L}_0} )\big]\ . \eq{5.6}
$$
After the integration  over the energy, momenta, Poisson resummation and
introduction
of  two auxiliary variables $\l, \bar \l$
(in order to `split' the $O(J^2)$ terms in ${\hat L_0}$, ${\hat
{\td L}_0} $ to be able to compute the trace over the  oscillators)
one finds$^{16}$
$$
 Z(r, \a,\b , q)  =   \int [d^2\t]_1 \ W(r, \a,\b,q|\t, \bar \t) \  , \eq{5.7}
$$  $$
\ \ \   [d^2\t]_1 \equiv
 {d^2\t \ \tau_2^{-14} }
 e^{4\pi \t_2} |f_0(e^{2\pi i \t})|^{-48}\  ,   $$
where $W$ is given  by the sum over windings and  the two auxiliary ordinary
integrals
$$ W =  {r \ov \a'\a\b\t_2 }
\sum_{w,w'=-\infty}^{\infty}
 \int d\l  d\bar \l \  e^{- {\cal I} (\h,\td\h,w,w',\tau,\bar \tau)}
\ {  \h \td \h |\theta'_1(0|\t )|^2 \ov \theta_1(\h|\tau )
\theta_1 (\td\h |\bar \t ) }   \ , \eq{5.8}  $$
$$
{\cal I} =
 {\pi\ov \a' \a\b\t_2 }
 \big[ \h \td \h   + \sqrt {\a'} r(q + \b )  (w'-  \t w) \td \h
 + \sqrt {\a'} r (  q -\a )
    (w'- \bar \t w)\h  $$
$$ +\  \a' r^2 q (q + \b -\a)  (w'-  \t w)(w'- \bar \t w)  + \ha \a'\a\b
 (\h - \td \h)^2 \big]\ , $$
$$ \h\equiv - \sqrt{\a'}[ 2  \b \l + q  r (w'-\t w) ] \ ,\ \  \ \
\td \h \equiv - \sqrt{\a'}[ 2 \a\bar  \l   + q  r (w'-\bar\t w)] \ .
 $$
 $Z$ in (5.7),(5.8) can be also obtained by  directly
computing the  string path integral with the action (3.5).
The special
 `null'  structure of (3.5) makes possible to compute this non-gaussian path
integral exactly (up to the two remaining ordinary  integrals over $\l, \ \bar
\l$).

Like the measure in (5.7)  $W$  is $SL(2,Z)$ modular invariant
(to show this one needs to shift $w,w'$ and redefine $\h, \td \h$).
$Z (r, \a,\b,q)$  has several
symmetry properties:
$Z(r,\a,\b,q) = $ $  Z(r,-\b,-\a ,q)=$ $ Z(r,-\a,-\b,-q) $ $ =
 Z(r,\b ,\a,-q).$
It  is also invariant under the  duality in $y$ direction
which transforms the theory with  $y$-period  $2\pi R$ and  parameters
$(\a,\b,q)$ into the theory  with
$y$-period  $2\pi \a'/R$  and  parameters $(q,\b-\a+q, \a)$ or
parameters  $(\a-\b-q,-q,-\b)$, i.e.
 $$
 Z(r,\a,\b,q) = Z(r\inv, q, \b -\a + q ,\a)   =
Z(r\inv, \a -\b - q , - q  ,-\b) .  $$
For  $\a=q$   or $\b=-q$
these  relations  take  their  standard  `circle' form:
$Z(r,\a,\b,\a)$ $  = Z(r\inv, \a, \b , \a) , $ $  \
 Z(r,\a,\b,-\b) = Z(r\inv, \a, \b ,-\b).$
When $\a=\b=q=0$ the partition function  (5.7) is  that of  the  free string
compactified on a circle of radius $R= \sqrt {\a'} r$.
Taking  the limit of the  non-compact $y$-dimension  ($R\to
\infty$) for generic $(\a,\b,q)$  one finds that
$Z$  (5.7),(5.8)
reduces  to the partition function of the free  bosonic closed string
theory.\footnote{$^l  $}{\baselineskip6pt This generalizes a similar
observation for the $\a=q=0$ model$^{14}$. In
the limit $R=\infty$ the
$\a=q=0$ model (1.1)  is equivalent to the model of ref.27
which has  trivial (free) partition function$^{31}$.}

The   expression for $Z$    simplifies
when at least
one of the   parameters  $\a, \b, q$ or $ q +\b-\a $
 vanishes so that the integrals over $\l, \bar  \l$ can be computed explicitly.
 For example, in the case when either $\a$ or $\b$ is equal to zero
(which includes the constant magnetic field model and $a=\sqrt 3 $ Melvin
model)
one finds
$$
 W=
  r  \sum_{w,w'=-\infty}^\infty    \exp\big(- {\pi \ov \t_2}
[r^2  (w'-\t w)(w'-\bar \t w) + \ha (\h _0- \td \h_0)^2]\big)
 {  \h_0 \td \h_0 |\theta'_1(0|\t )|^2 \ov
\theta_1(\h_0|\tau )
\theta_1 (\td\h_0 |\bar \t ) }  , $$
$$
\h_0  = \sqrt {\a'} (q + \b )  r(w'-\t w) \ , \ \ \ \
\td \h_0 =\sqrt {\a'} (q - \a) r(w'-\bar \t w) \ , \ \  \ \a\b=0 \ .    $$
The  magnetic instability of these models  (the presence of
tachyons  in the spectrum)
is reflected  in  singularities (or imaginary parts) of $Z$.
The partition function has  new divergences at  critical values of the magnetic
field  parameters  when  the energy develops an imaginary
part.
\vglue0.6cm
\leftline{\tenbf 6. Concluding remarks }
\vglue0.4cm
The actions (1.1)--(1.3),(3.5)
admit straightforward $(1,1)$ and $(0,1)$ (or $(1,0)$)  supersymmetric
generalizations
describing  closed  superstring and heterotic string
models where the two  abelian   magnetic fields
appear in the Kaluza-Klein sector.
In addition, it is possible to construct the heterotic string
versions of (1.1) and (1.2) which correspond to the same background fields
(2.8) and (2.9)
 but  now the magnetic field\footnote{$^m  $}{\baselineskip6pt
Note that the two vector fields  $\cal A$ and $\cal B$
in (2.8),(2.9) are the same up to sign.}
 is embedded in the
gauge sector of the heterotic string.$^{14,15}$
The idea is to `fermionize' the internal bosonic coordinate $y$.
The non-trivial part of the action of the
resulting   heterotic string analog of
(1.1),(2.8)
is$^{14}$
$$ I_{(0,1)} = {  {1\ov \pi \a'} }
\int d^2 \s \big[ -\del t \bd t    -  2A_i \del t  \bd x^i  +
(\delta_{ij} - A_iA_j) \del x^i   \bd  x^j $$  $$
 - \   \l_{L}^\htt \del \l^\htt_L +   \l_{Li}\del \l^i_L
  + F_{ij} \del  t    \l_{L}^i\l^j_L  +  \ha F_{ij}  A_k  \del  x^k
\l_{L}^i\l^j_L $$ $$
 +\   \bar \psi_{R} ( \bd -  ie_0 A_i \bd x^i) \psi_{R}
+  \ha i e_0  F_{ij}   \bar \psi_{R} \psi_{R}\l_L^i\l_L^j\ \big]
\   ,  \    \  \   e_0\equiv R\inv = {  \sqrt {2/\a'}}\   ,   $$
where $A_i = -\ha \b \ep_{ij}x^j$.
Like  the bosonic model (1.1) and its direct supersymmetrisations
 this model can be solved explicitly$^{41}$.
Detailed study of the pattern of the spectrum  in this and similar models
(in particular, the  special cases when certain higher spin states
become massless before becoming tachyonic)
may teach us about possible hidden string symmetries.
Another interesting direction  seems  to
consider these models in the context of  electro-magnetic duality.

Let us mention  also that there  was a suggestion$^{19,21,42}$
to interpret  the Melvin-type solutions of the  higher dimensional
 (dilaton) Einstein-Maxwell  theory
as alternatives to the standard Kaluza-Klein compactification on compact
spaces.
The idea was to consider  the $(\r,\vp)$ part of the Melvin space
(cf.(2.9),(2.10))  as an
internal one. Though this 2-space is non-compact, it is `nearly
closed' and   the corresponding
 scalar Laplacian    has  discrete branch in the spectrum (cf.(5.3),(5.2)).
The conformal  Melvin models  (1.2),(1.3)
  may be  used in  an  attempt of  string-theoric
implementation of this  idea of  having   a non-compact space
as an internal one
(in string  models (1.2),(1.3)  the internal space  is  3-dimensional
$(\r, \vp, y)$).
Since the spectrum of the string mass operator for the Melvin model
is explicitly computable, this makes possible to determine the
corresponding masses of particles  moving in  extra flat spatial  dimentions.
Unfortunately, as in the case of the particle theory limit$^{42}$,
this idea does not actually work in   the Melvin model:
though most of the states in the spectrum belong to its discrete  branch,
there  are also  special ``zero mode"
 states (e.g.,
scalar state with zero charge and orbital momentum in (5.2))  which  have
continuous mass parameter.
It may happen, however, that there are related string models
which (like  modifications of the Melvin solution  discussed in ref.{42}) may
not have this deficiency and yet  be explicitly
{\it solvable}. Such models  could  be  also
of interest in the context of possible supersymmetry breaking by magnetic
backgrounds in internal dimensions
(see ref.43 and refs. there).
 \vglue0.6cm
\leftline{\tenbf 7. Acknowledgements}
\vglue0.4cm
I would like to  acknowledge  useful discussions  and collaboration  with J.
Russo
and   the support of PPARC,   EEC grant SC1$^*$-CT92-0789 and NATO grant CRG
940870.


\gdef \jnl#1, #2, #3, 1#4#5#6{ { #1~}{\bf #2} (1#4#5#6) #3}

\vglue0.6cm
\leftline{\tenbf 8. References}
\vglue0.4cm

\itemitem{1.} L.D. Landau, {\it Z. Phys.} {\bf 64 }
 (1930) 629.

\itemitem{2.} J. Schwinger, \pr {\bf 82  }
(1951) 664.

\itemitem{3.} A. Salam  and J. Strathdee, \np {\bf B90  }
(1975) 203; A. Linde, {\it Rep. Progr. Phys. } {\bf 42 }
 (1979) 389.

\itemitem{4.} G.K. Savvidi, \pl {\bf B71  }
(1977) 133.

\itemitem{5.} N.K. Nielsen and P. Olesen,
\np {\bf B144 }
 (1978) 376;
V.V. Skalozub, {\it Sov. J. Nucl. Phys.} {\bf 23 }
 (1978) 113;
H. Leutwyler, \np {\bf B179 }
 (1981) 129.

\itemitem{6.} J. Ambjorn and P. Olesen, \np {\bf B315 }
 (1989) 606; {\bf B330 }
 (1990) 193.

\itemitem{7.} E.S. Fradkin and A.A. Tseytlin, \pl {\bf
B163  }
(1985) 123.

\itemitem{8.} A. Abouelsaood, C. Callan, C. Nappi and S. Yost, \np {\bf B280 }
 (1987) 599.

\itemitem{9.} C.P. Burgess, \jnl \np, B294, 427,  1987; V.V. Nesterenko, \jnl
\ijmp,   A4, 2627,  1989.

\itemitem{10.} S. Ferrara and M. Porrati, \mpl {\bf A8 }
 (1993) 2497.

\itemitem{11.}  P.C. Argyres and C.R. Nappi, \pl {\bf B224} (1989) 89;
S. Ferrara, M. Porrati and V.L. Telegdi, Phys. Rev. {\bf D46 }
(1992) 3529.

\itemitem{12.} E.S. Fradkin and A.A. Tseytlin, \np {\bf B227 }
 (1983) 252.

\itemitem{13.} J.G. Russo and L. Susskind,   \np {\bf B437  }
(1995) 611; J.G. Russo, \pl {\bf B335  }
(1994) 168; A. Sen, \np {\bf B440} (1995) 421.

\itemitem{14.} J.G.  Russo and A.A. Tseytlin, ``Constant magnetic field in
closed
string theory: an exactly solvable model", CERN-TH.7494/94,
 hep-th/9411099.

\itemitem{15.} A.A. Tseytlin, \jnl \pl, B346, 55, 1995.

\itemitem{16.} J.G. Russo and A.A. Tseytlin,  ``Exactly solvable string models
of curved space-time backgrounds",
CERN-TH/95-20, hep-th/9502038.

\itemitem{17.} M.A. Melvin, \pl{\bf  8 }
 (1964) 65; W.B. Bonner, {\it Proc. Phys. Soc. London}
{\bf A67 }
 (1954) 225.

\itemitem{18.} A. Abrikosov, {\it  Zh. Exp. Teor. Fiz. (JETP)} {\bf 32} (1957)
1441;
H.B. Nielsen and P. Olesen, \np {\bf B61 }
 (1973) 45.

\itemitem{19.} G.W.  Gibbons, in: {\it Fields and Geometry}, Proceedings of the
22-nd Karpacz
Winter School of Theoretical Physics, ed. A. Jadczyk (World Scientific,
Singapore,  1986).

\itemitem{20.} H.F. Dowker, J.P. Gauntlett, D.A. Kastor and J. Traschen,
\pr {\bf D49 }
 (1994) 2909;
 H.F. Dowker, J.P. Gauntlett, S.B. Giddings and G.T. Horowitz, \pr {\bf D50  }
(1994)
2662.

\itemitem{21.} G.W.  Gibbons and  K. Maeda, \np {\bf B298 }
 (1988) 741.

\itemitem{22.}  G.H.  Horowitz and A.A. Tseytlin,  \pr  {\bf D51  }
 (1995) 2896; \pr {\bf D50 }
 (1994) 5204.

\itemitem{23.}   E. Kiritsis and C. Kounnas,  ``Curved four-dimensional
spacetime
as infrared regulator in superstring theories", hep-th/9410212;
``Infrared regularization
of superstring theory and the one-loop calculation of coupling constants",
hep-th/9501020.

\itemitem{24.}  D.A. Lowe and A. Strominger, \prl {\bf 73  }
(1994) 1468.

\itemitem{25.} S.  Giddings, J. Polchinski and A. Strominger,   \pr {\bf D48 }
 (1993)
 5784.

\itemitem{26.}  T. Banks, M. Dine, H. Dijkstra and W. Fischler, \pl {\bf B212
}
(1988) 45;
I. Antoniadis, C. Bachas and A. Sagnotti, \pl {\bf B235  }
(1990) 255;
R. Khuri, \pl{\bf  B259  }
(1991) 261; \np {\bf B387 }
 (1992) 315;
Gauntlett, J. Harvey and J. Liu,
\jnl \np, B409, 363, 1993;
C. Bachas and E. Kiritsis, \pl {\bf B325 }
 (1994) 103.

\itemitem{27.}
 C. Nappi and E. Witten, \jnl \pl,  B293, 309, 1992.

\itemitem{28.}
 I. Bars, in: {\it  ``Perspectives in mathematical physics"}, vol.3, eds. R.
Penner and  S.-T. Yau
(International Press, 1994); hep-th/9309042.

\itemitem{29.}
C. Kounnas, in: {\it  Proceedings of the International
Europhysics Conference on High Energy Physics}, Marseille, 22-28 July,  1993;
hep-th/9402080.

\itemitem{30.}
 D.  Olive,
 E. Rabinovici and A. Schwimmer,  \jnl \pl, B321, 361, 1994.

\itemitem{31.}
E. Kiritsis and C. Kounnas, \jnl \pl, B320, 264, 1994;
E. Kiritsis, C. Kounnas and  D. L\"ust, \jnl \pl,  B331, 321,  1994.

\itemitem{32.}
K. Sfetsos,  \jnl \pl, B324, 335, 1994;
\ijmp {\bf  A9 }
 (1994) 4759;
\jnl \pr,  D50, 2784,  1994.

 \itemitem{33.}
N. Mohammedi, \jnl \pl,  B325, 379,  1994;
 J.M. Figueroa-O'Farrill and S. Stanciu, \jnl \pl, B327, 40,  1994.

\itemitem{34.}
I. Antoniadis and N. Obers, \jnl \np, B423, 639, 1994;
K. Sfetsos and A.A. Tseytlin, \jnl \np, B427, 245, 1994.

\itemitem{35.} J. Maharana and J. Schwarz, \jnl \np, B390, 3, 1993.

\itemitem{36.} P. Forg\' acs, P.A. Horv\' athy, Z. Horv\' ath and L. Palla,
``The Nappi-Witten string in the light-cone gauge", hep-th/9503222.

\itemitem{37.} J. Horne and G.T. Horowitz, \np {\bf B368} (1992) 444.

\itemitem{38.} M. Ro\v cek and E. Verlinde, \jnl \np, B373, 630, 1992.

\itemitem{39.}  R. Dijkgraaf, H. Verlinde and E. Verlinde, \np {\bf B371  }
(1992) 269;
E. Kiritsis, \jnl \np, B405, 109, 1993.

\itemitem{40.} A. Dabholkar, \jnl \np, B439, 650, 1995, hep-th/9408098.

\itemitem{41.} J.G. Russo and A.A. Tseytlin, to appear.

\itemitem{42.}  G.W.  Gibbons and  D.L.  Wiltshire, \np {\bf B287} (1987) 717.

\itemitem{43.} C. Bachas, ``A way to break supersymmetry", hep-th/9503030.

\vfill\eject

\bye